\documentclass[preprint,12pt]{elsarticle}

\usepackage{amssymb}
\usepackage{color}
\usepackage{newtxtext,newtxmath}

\usepackage{pdflscape}
\usepackage{fancyhdr} 

\fancypagestyle{mylandscape}{
\fancyhf{} 
\fancyfoot{
\makebox[\textwidth][r]{
  \rlap{\hspace{1.20cm}
    \smash{
      \raisebox{4.00in}{
        \rotatebox{90}{\thepage}}}}}}
}

\journal{International Journal of heat and fluid flow}

\begin{document}

\begin{frontmatter}



\title{Euler-Lagrange study of Microbubble-Laden Turbulent Flow over Superhydrophobic surfaces}


\author[1]{Byeong-Cheon Kim}
\author[1]{Kyoungsik Chang}
\author[1]{Sang-Wook Lee}   

\affiliation[1]{organization={School of Mechanical Engineering},
            addressline={University of Ulsan}, 
            city={Ulsan},
            country={Republic of Korea}}
 
\author[2]{Jaiyoung Ryu}
\affiliation[2]{organization={Department of Mechanical Engineering},
            addressline={Korea University}, 
            city={Seoul},
            country={Republic of Korea}}
            
\author[3]{Minjae Kim}
\author[3]{Jaemoon Yoon}
\affiliation[3]{
            addressline={5th R\&D Institute, Agency for Defense Development}, 
            city={Changwon},
            country={Republic of Korea}}

\begin{abstract}
For slow-speed ships, underwater vehicles, and pipe transportation systems, viscous resistance accounts for a large proportion of the total energy losses. As such, various technologies have been developed to reduce viscous resistance and enhance energy efficiency in these applications. Air injection and surface treatment are two representative drag reduction techniques. Additionally, efforts to combine multiple drag-reduction techniques have been the subject of extensive research. In this study, the synergistic effects of integrating microbubble injection and superhydrophobic Surface(SHS) drag reduction approaches were analyzed. A 2-way coupling Euler-Lagrange approach was used alongside direct numerical simulation, based on the spectral element method, to investigate the synergistic effects of applying two separate drag reduction methods. Three types of SHS were investigated in our simulations; post type, transverse ridge type, and ridge type. The drag reduction performances and flow characteristics of the various configurations, with and without microbubble injection, were compared in a turbulent horizontal channel flow with $Re_{\tau}=180$. The results of these tests showed that, combining post-type SHS with microbubbles was the most effective, producing a synergistic drag reduction effect. However, combining microbubble injection with ridge-type SHS increased drag relative to ridge-type SHS alone, showing the importance of carefully selecting wall type for the best possible performance.
\end{abstract}


\begin{highlights}
\item 2-way coupling Euler-Lagrange approach based on spectral element method; Nek5000.
\item Sinergistic drag reduction of microbubble-laden turbulent flow over superhydrophobic surface.
\item Turbulence modulation by microbubbles and wall type.
\end{highlights}


\end{frontmatter}


\section{Introduction}
\label{sec:Introduction}

With the increasingly stringent regulations on greenhouse gas emissions, the development of energy-efficient technologies has become a critical focus in various industries including shipbuilding and marine engineering. In these fields, the frictional drag on slow-speed ships and underwater vehicles caused by water resistance accounts for a significant portion of their total energy losses\cite{fukuda2000frictional}. Consequently, significant research efforts have been directed at mitigating frictional drag to reduce fuel consumption and enhance the overall energy efficiency of these vehicles.

Techniques for reducing viscous resistance can be categorized into two primary approaches; active and passive approaches\cite{wang2022drag}. Active approaches involve the injection of additives into boundary flow, these additives include polymers\cite{white2008mechanics,marchioli2021drag} and, air\cite{ceccio2010friction, murai2014frictional}. Notably, the air injection methods are distinguished by the gas flow rate used, they can also be unclassified into air layer drag reduction(ALDR), bubble drag reduction(BDR), and microbubble drag reduction(MBDR)\cite{murai2014frictional}. On the other hand, passive approaches involve surface treatments that produce micro-structures on the solid surface in the flow, these include superhydrophobic surface(SHS)\cite{rothstein2010slip,lee2016superhydrophobic,park2021superhydrophobic} and riblet\cite{dean2010shark}. 

Each approach has its advantages and disadvantages. The active approach has the advantage of directly producing a drag reduction effect. However, it has the drawback of the additives being difficult to control\cite{jang2014experimental,tanaka2021repetitive}, it also requires continuous energy consumption to maintain reduced drag. The passive approach has the benefit of not requiring additional energy, but at high Reynolds numbers, the plastron within the superhydrophobic structures can be lost due to turbulence, leading to a decrease in its drag reduction performance.

Recently, several investigations on combining active and passive approaches to maximize drag reduction have been conducted. This integrated strategy aims to leverage the advantages of both methods to enhance overall performance and energy efficiency in various applications. An approach where air injection is combined with SHS to compensate for the loss of plastron (air) within superhydrophobic structures in high Reynolds number flows has been proposed. Injecting air maintains a stable air layer on the surface within the flow, preventing air loss and ensuring a sustained drag reduction effect\cite{breveleri2023plastron}. Zhu et al.\cite{zhu2024achieving} conducted experiments regarding the synergistic drag reduction effect achieved by combining randomly distributed SHS and air injection. The integration of SHS with microbubble air injection has considerable potential to enhance and maintain drag reduction effects. However, studies and practical applications of this synergistic mechanism still need to be explored.

In this study, we investigate the synergistic drag reduction effects of microbubble injection over SHS using direct numerical simulation (DNS), based on the spectral element method (SEM) Nek5000 code, and a two-way coupling Euler-Lagrange approach. A horizontal channel configuration was adopted to investigate the drag reduction effects of microbubble injection over SHS. SHS were applied on one-side of a channel while the shear Reynolds number($Re_{\tau}$) of the flow was set to 180 with a constant pressure gradient. Four different types of wall conditions were considered; no-slip condition, post-type SHS, transverse ridge-type SHS, and ridge-type SHS. This research aims to deepen our understanding of the synergistic interaction between microbubbles and SHS while also providing insights to help optimize practical drag reduction strategies for applications.

\section{Numerical method}
\label{sec:Numerical Methods}
The DNS based on the Euler-Lagrange approach is used to resolve microbubble-laden turbulent flow. The approach considers two phases: the continuous phase (the carrier fluid that is present in bulk) and dispersed phase (usually in the form of rigid solid particles, deformable drops, or bubbles). In this study, water is the continuous phase and air bubbles are the dispersed phase. Section 2.1 presents the governing equation of the continuous phase and describes features of the Nek5000 code used in this work. In Section 2.2, the governing equation for the dispersed phase is explained. Finally, Section 2.3 presents the modeling method for the interaction between the continuous and dispersed phases.

\subsection{Continuous phase (water)}
\label{subsec:Cont. phase}
Nek5000 \cite{fischer2008nek5000} is an open-source CFD solver that utilizes SEM and employs a high-order polynomial function as the basis function. The solver exhibits minimal numerical dissipation and dispersion errors while achieving efficient parallelization\cite{patera1984spectral}.
The continuous phase is considered to be an incompressible Newtonian fluid, which is resolved using SEM. The non-dimensional governing equations for the continuous phase are described by Eqs.(\ref{eqn:ge1})-(\ref{eqn:ge3}).
\newline
\begin{equation}\label{eqn:ge1}
\nabla\cdot{\textbf{u}}^*=0
\end{equation}

\begin{equation}\label{eqn:ge2}
\frac{\partial \textbf{u}^*}{\partial t^*} + \textbf{u}^* \cdot \nabla \textbf{u}^* = -\nabla p^*+ \frac{1}{Re} \nabla \cdot \boldsymbol{\tau}^*+ \textbf{f}^*_{PG} + \textbf{f}^*_i
\end{equation}

\begin{equation}\label{eqn:ge3}
\boldsymbol{\tau}^*=[\nabla \textbf{u}^* + \nabla \textbf{u}^{*\top}]
\end{equation}
\newline
Equation (\ref{eqn:ge1}) is a non-dimensional continuity equation while equation (\ref{eqn:ge2}) is a non-dimensional Naver-Stokes equation. All the parameters are normalized by the channel half height $h$ as the reference length and the bulk velocity $U_b$ as the reference velocity. The normalized parameters are expressed with superscript *. The variables $u^*$ and $p^*$ are the velocity vector and pressure respectively. $Re$ is the bulk Reynolds number expressed as $Re=\rho_l U_b h/\mu_l$. The parameters $\rho_l$ and $\mu_l$ are the respective density and viscosity of water. The subscript $l$ indicates these parameters relate to the continuous phase. $f_{PG}^*$ gives the constant pressure gradient along the channel. $f_i^*$ is a momentum interaction term that relates the continuous phase and dispersed phase. The momentum interaction term is calculated according to the standard Euler-Lagrange formulation given in previous research\cite{asiagbe2017large,elgobashi2006updated,pang2014numerical,velasco2022numerical,zhai2020simulation}.

\subsection{Dispersed phase (air)}
\label{subsec:Disp phase}
The Lagrange particle tracking approach is used to predict the microbubble dynamics. This approach is integrated with the Nek5000 solver, and provides one-way and two-way coupling options. The microbubbles are assumed to be point particles with a rigid spherical shape, they are tracked by Newton’s second law of motion in a turbulent flow field\cite{maxey1983equation}. Therefore, each microbubble is tracked by the equation:
\newline
 \begin{eqnarray}\label{eqn:bubble eqn}
    \rho^*_b \frac{d\textbf{u}^*_b}{dt^*} =\frac{3C_D}{4d^*_b} \left|\textbf{u}^*_l-\textbf{u}^*_b \right|(\textbf{u}^*_l-\textbf{u}^*_b)+C_{VM}(\frac{D\textbf{u}^*_l}{Dt^*}-\frac{d\textbf{u}^*_b}{d t^*})\nonumber 
    \\ 
    +\frac{D\textbf{u}^*_l}{D t^*}+(1-\rho^*_b)a^*_g+C_L(\textbf{u}^*_l-\textbf{u}^*_b)\times\boldsymbol{\omega}^*_l
 \end{eqnarray}
\newline

$\rho_b^*$ is the bubble density normalized according to the liquid density($\rho_b^* = \rho_b / \rho_l$) while the subscript $b$ indicates these parameters relate to the dispersed phase. The Naumann-Schiller model\cite{schiller1933drag} is usually used to calculate the drag coefficient of rigid spherical particles. 

However, this model is not suitable for resolving the rising velocity of bubbles. Motivated by this, Tomiyama\cite{tomiyama2002terminal} proposed a drag model for these bubbles. This model has achieved good agreement with the experimental values\cite{duineveld1995rise}. 
\newline
\begin{equation}
    C_D=max\left(min\left(\frac{16}{Re_b}(1+0.15Re_b^{0.687}),\frac{48}{Re_b}\right),\frac{8}{3}\frac{Eo}{Eo+4}\right)
\end{equation}
\begin{equation}
    Re_b=\frac{\rho_l\left|\textbf{u}^*_l-\textbf{u}^*_b \right|d^*_b}{\mu_l}
\end{equation}
\newline
The drag coefficient correlation is a function of the bubble Reynolds number, also known as particle Reynolds number. The virtual mass coefficient $C_{VM}$ is 0.5 for these rigid spherical objects\cite{brennen1982review}. The pressure gradient force is considered to model the pressure difference between the bubbles in the continuous phase, and the expression is adopted from Maxey and Riley\cite{maxey1983equation}. $a^*_g$ represents the non-dimensional acceleration due to gravity. 
The lift coefficient $C_L$ was derived by Legendre and Magnaudet\cite{legendre1998lift}. In this lift model, $C_L$ is a function of the high Reynolds number lift coefficient($C_L^{highRe}$) and low Reynolds number lift coefficient($C_L^{lowRe}$).
\newline
\begin{equation}
    C_L=\sqrt{{C_L^{lowRe}}^2+{C_L^{highRe}}^2}
\end{equation}
\begin{equation}
    C_L^{lowRe}(Re_b,Sr_b)=\frac{6}{{\pi}^2{(Re_bSr_b)}^{0.5}}\left[\frac{2.255}{{(1+0.2\zeta^{-2})}^{1.5}} \right]
\end{equation}
\begin{equation}
    C_L^{highRe}(Re_b)=\frac{1}{2}\frac{Re_b+16}{Re_b+29}
\end{equation}
\begin{equation}
    Sr_b=\frac{\left|\omega_b\right|d_b}{2\left|u^*_l-u^*_b\right|}, \zeta=\sqrt{\frac{Sr_b}{Re_b}}
\end{equation}
\newline
In the above equation, $Sr_b$ is the non-dimensional shear rate. $\omega_l^*$ is the fluid vorticity at the bubble location, which is written as $\omega_l^*=0.5\times\nabla u$.

\subsection{Phase interaction(two-way coupling)}
The force which acts on the continuous phase from the dispersed phase is considered as two-way coupling Euler-Lagrange framework. When the dispersed phase volume fraction $\alpha$ is higher than $10^{-6}$, the fluid-bubble interaction affects the turbulent structure of the continuous phase\cite{elgobashi2006updated}. In this situation, the momentum equation coupling between the two phases is needed by the momentum interaction term, $f_i^*$. 
\newline
\begin{equation}
    f_i^*(x^*)=-\frac{1}{V}\sum_{i=1}^{N_b}F^*_i(x^*)g_M(\left|x^*-X^*\right|)
\end{equation}
\newline
Here, the summation is conducted over the total number of bubbles, $N_b$ in the cell volume. According to Newton’s third law, the negative sign is multiplied on the hydrodynamic force term $F_i^*$. The drag force, shear-lift, pressure gradient force, and virtual mass force are included in the particle-fluid force $F_i^*$. $x^*$ and $X^*$ are non-dimensional coordinates of the continuous phase and dispersed phase, respectively. The kernel for the Gaussian filter $g_M$, is written as 
\newline
\begin{equation}
    g_M(\left|x^*-X^*\right|)=\frac{1}{{(\sigma\sqrt{2\pi})}^3}exp\left[-\frac{{(\left|x^*-X^*\right|)}^2}{2{\sigma}^2}\right]
\end{equation}

\section{Computational Setup}

\begin{figure}
    \centering
    \includegraphics[width=1.0\textwidth]{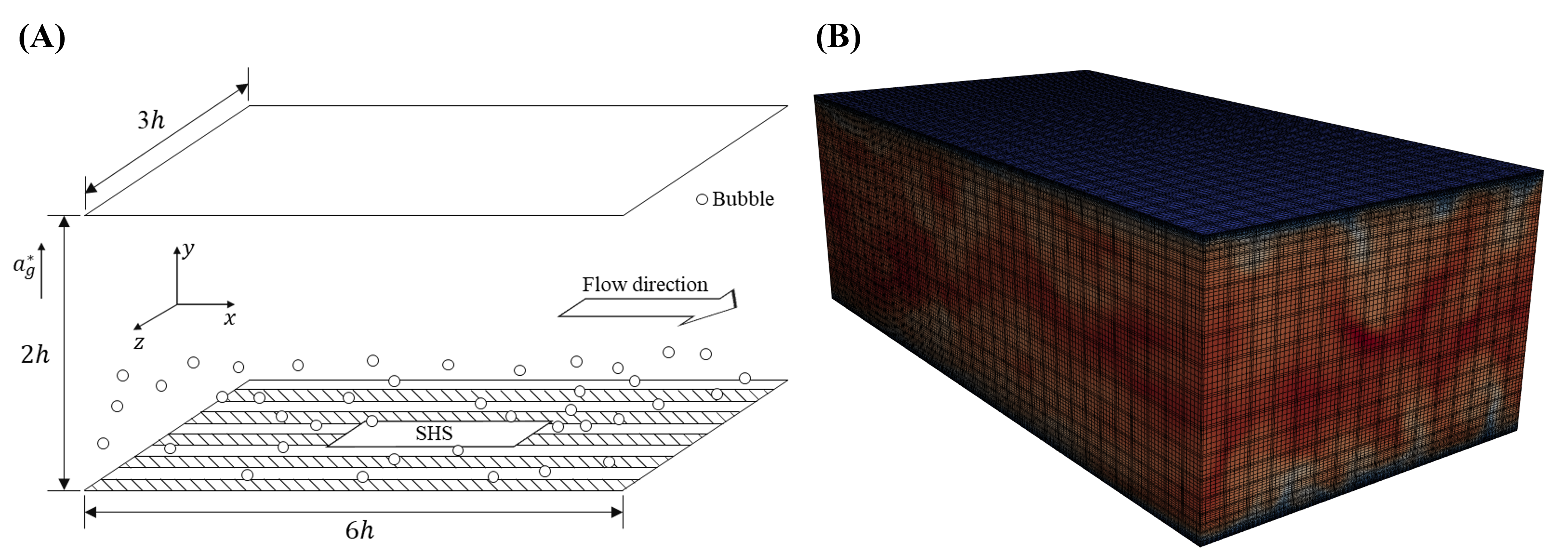}
    \caption{Schematic diagram of the computational domain and mesh configuration for microbubble-laden turbulent flow over a superhydrophobic surface (SHS). (A) Schematic representation of the computational domain with microbubble-laden flow over SHS. (B) 3D visualization of the computational mesh used for the direct numerical simulation (DNS).}
    \label{fig:geo1}
\end{figure}

The simulation domain consists of a horizontal channel composed of two parallel walls. Figure. \ref{fig:geo1} (A) shows a schematic diagram of the horizontal channel with microbubbles and SHS. The domain size$(L_x\times L_y\times L_z)$ is \textit{6h$\times$2h$\times$3h}, as shown in Figure. \ref{fig:geo1}. (B) shows a 3D visualization of the computational mesh and domain. The simulation domain was divided into hexahedral elements and discretized into \textit{32$\times$24$\times$16} macro mesh elements. Each element is divided by the polynomial order (\textit{$N_p$}) into Gauss-Lobatto-Legendre(GLL) grid points. A 7th order polynomial was used in this work. This produces about 4.2 million GLL grid point. The elements are uniformly distributed in streamwise and spanwise directions. The average grid spacing in the streamwise direction $\Delta x^+$ is 4.8 wall units and in the spanwise direction $\Delta z^+$ is 4.8 wall units. The grid spacing increases as we move away from the wall, this increased density near the wall helps to resolve the smallest scale vortices in the normal direction. The grid spacing near the wall in the normal direction $\Delta y_w^+$ is about 0.089 and while the largest grid spacing in the same direction $\Delta y_c^+$ is 7.65 near the center of the channel. This grid resolution is comparable to those found in previous literature for turbulent channel DNS using Nek5000\cite{zhai2020simulation, mortimer2019near}. 

\begin{figure}
    \centering
    \centerline{\includegraphics[width=1.4\textwidth]{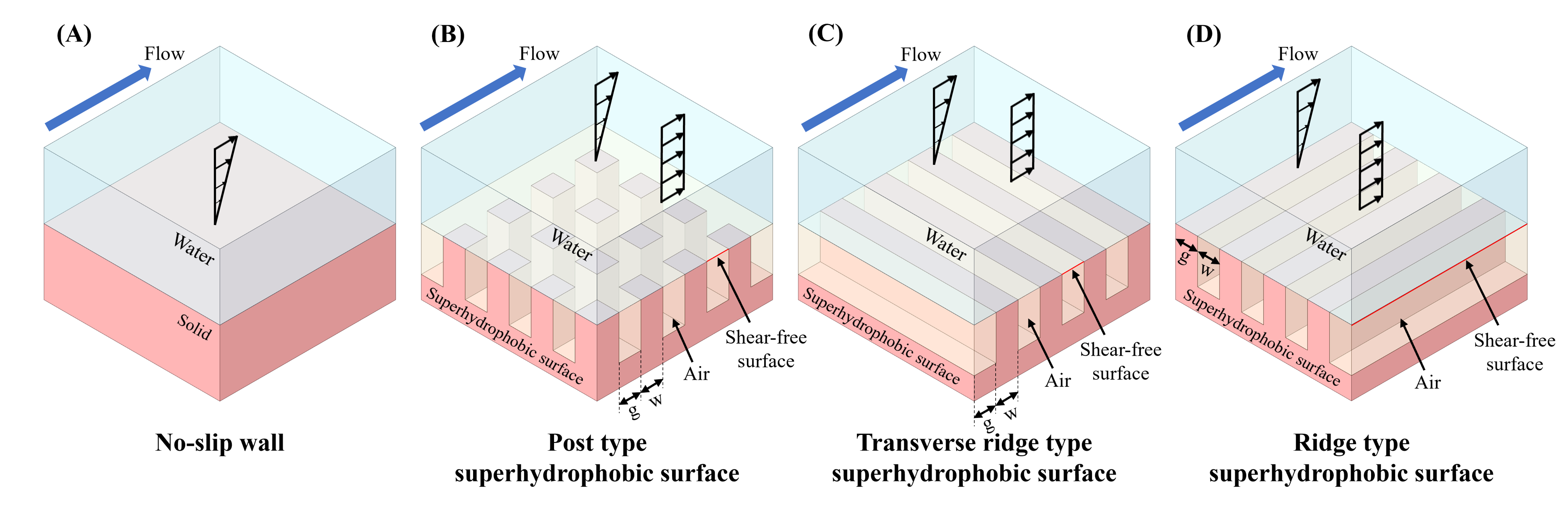}}
    \caption{Schematic diagram of the various wall conditions with either no-slip condition or boundary conditions based on the wall's superhydrophobic structure.
            (A) No-slip wall, (B) Post type, (C) Transverse ridge type, (D) Ridge type}
    \label{fig:SHS}
\end{figure}

In the continuous phase, a no-slip wall condition was set for the top surface. An SHS boundary condition was set for the bottom surface like the first used by Min and Kim\cite{min2004effects,min2005effects}. This SHS boundary condition is composed of either a no-slip wall condition or no-shear stress condition according to the part of the structure water is flowing over. Three SHS boundary condition types were considered: post type, transverse ridge type, and ridge type\cite{martell2009direct,martell2010analysis}. 
Figure. \ref{fig:SHS} is a schematic diagram showing the different wall types used. Figure. \ref{fig:SHS} (B,C,D) show schematic diagrams of each type of SHS structure used along with their associated boundary conditions. Walls with these SHS structures have two different interfaces with the water flowing over them : a no-slip surface, and a shear-free surface which is an interface between water and air. Several post and ridge configurations were adopted, in all configurations the micro-feature width $(w^*)$ and feature gap spacing $(g^*)$ were set as 0.1875$h$. Therefore, the post-type SHS consists of eight lines of posts in the spanwise direction. The ridge-type SHS has eight ridges in the spanwise direction, while the transverse-ridge-type SHS has sixteen ridges in the streamwise direction. The details of these configuration are shown in Table. \ref{table:case descriptions}.

For the dispersed phase, an elastic reflective collision condition was applied. If the microbubble center comes within the bubble radius of the wall, a collision between the microbubble and wall is considered to have occured and the sign of the microbubble’s normal velocity is swapped\cite{asiagbe2017large,pang2014numerical,zhai2020simulation}. Periodic boundary conditions are also applied in the streamwise and spanwise directions for both continuous and dispersed phase. Gravitational acceleration($a_g^*$) is applied in the positive $y$ direction for interactions between buoyant microbubbles and the SHS.

\begin{table}
\caption{The case descriptions and configuration details}
\centering\resizebox{\textwidth}{!}{
\begin{tabular}{c c c c c c}
 \hline
 Name & SHS type & micro-feature width, $w^*$ & micro-feature gap spacing, $g^*$ & Microbubble\\ \hline
 Ns & No-slip wall & 0.1875$h$ & 0.1875$h$ & -  \\
 NsMB & No-slip wall & 0.1875$h$ & 0.1875$h$ & O \\
 Po & Post type & 0.1875$h$ & 0.1875$h$ & - \\
 PoMB & Post type & 0.1875$h$ & 0.1875$h$ & O \\
 Tr & Transverse ridge type & 0.1875$h$ & 0.1875$h$ & - \\
 TrMB & Transverse ridge type & 0.1875$h$ & 0.1875$h$ & O \\
 Ri & Ridge type & 0.1875$h$ & 0.1875$h$ & - \\
 RiMB & Ridge type & 0.1875$h$ & 0.1875$h$ & O  \\ \hline
\end{tabular}}
\label{table:case descriptions}
\end{table}

Initially, the single-phase flow was driven by a streamwise constant pressure gradient corresponding with the desired Reynolds number. At a fully developed channel, the average wall shear stress $\tau_w$ is directly related to the average channel pressure gradient, $\partial P / \partial x$ by $\tau_w = (2/L_y )(\partial P /\partial x)$\cite{pope2000turbulent}. The frictional Reynolds number $Re_\tau$ was set at 180 as calculated by $Re_\tau=\rho_l u_\tau h/\mu_l$. Where $\rho_l$ and $\mu_l$ represent water density and dynamic viscosity, respectively. After the single-phase flow was fully developed, microbubbles are introduced into the simulation domain. The non-dimensional diameter of the microbubbles $d_b^*$ is set to $0.01h$ while the density of microbubbles $\rho_b^*$ is set to $1.3/1000$. The void fraction $\alpha$ is $1.12 \times 10^{-4}$ while the corresponding number of microbubbles $N_b$ to give this void fraction is $5,785$. 

The hypothesis that the microbubbles should be modeled as rigid spheres is derived using a Grace diagram\cite{grace1976shapes}, based on three non-dimensional numbers that govern bubble shape: the bubble Reynolds number ($Re_{b}$), the Eötvös number ($Eo$), and the Morton number ($Mo$). The Eötvös number, a measure of bubble deformability, represents the ratio of buoyancy to surface tension forces and is defined as:

\begin{equation}
    Eo=\frac{(\rho_l-\rho_b)d_b^2a_g}{\sigma_l}
\end{equation}
where $\sigma_l$ is the surface tension of the continuous phase fluid. When the Eötvös number is less than 0.2, it is reasonable to assume the bubbles will have a rigid spherical shape\cite{clift2005bubbles}. For water, $\sigma_l=0.0728 N/m$, and for microbubbles with diameters ranging from 1 to 1000 $\mu m$, the calculated $Eo$ ranges from $1.346\times 10^{-7}$ to $0.1346$. The Morton number, representing bubble deformability due to viscous effects, is defined as : 

\begin{equation}
    Mo=\frac{a_g \vert{\rho_l-\rho_b}\vert \mu_l^4}{\rho_l^3\sigma_l^3}
\end{equation}
For water, $Mo$ is $2.63\times10^{-11}$. According to the Grace diagram, bubbles within this range of $Eo$ and $Mo$ are classified as non-deformable, as such they maintain a spherical shape. Therefore, the assumption that microbubbles take the form of rigid spheres is valid.

The $P_N-P_N$ formulation is used for velocity and pressure with polynomial order $N_p$. An operator-integration-factor splitting (OIFS) scheme was adopted for temporal discretization. The linear terms are solved implicitly with 3rd-order backward differentiation(BDF3), while non-linear terms are solved using a characteristics-based explicit scheme. The OIFS method allows for larger time steps than would be allowed with regular explicit/extrapolation schemes \cite{fischer2003implementation}.
After calculation of the continuous phase, the locations and velocities of the microbubbles are interpolated at the center of the microbubbles where the spectral accuracy depends on the polynomial order, $N_p$ in Nek5000\cite{offermans2017gather}. With high-order Lagrange interpolation, the locations and velocities of the microbubbles can be more accurately resolved. Therefore, a low error can be achieved with a high polynomial order($N_p$)\cite{zwick2020scalable}. The positions of the microbubble are tracked by integration of the Lagrangian tracking equation into the 3rd-order Runge-Kutta scheme.

\section{Results}
\label{sec:Results}
\subsection{Validation}
\subsubsection{Tomiyama drag model}

\begin{figure}
    \centering
    \includegraphics[width=0.8\textwidth]{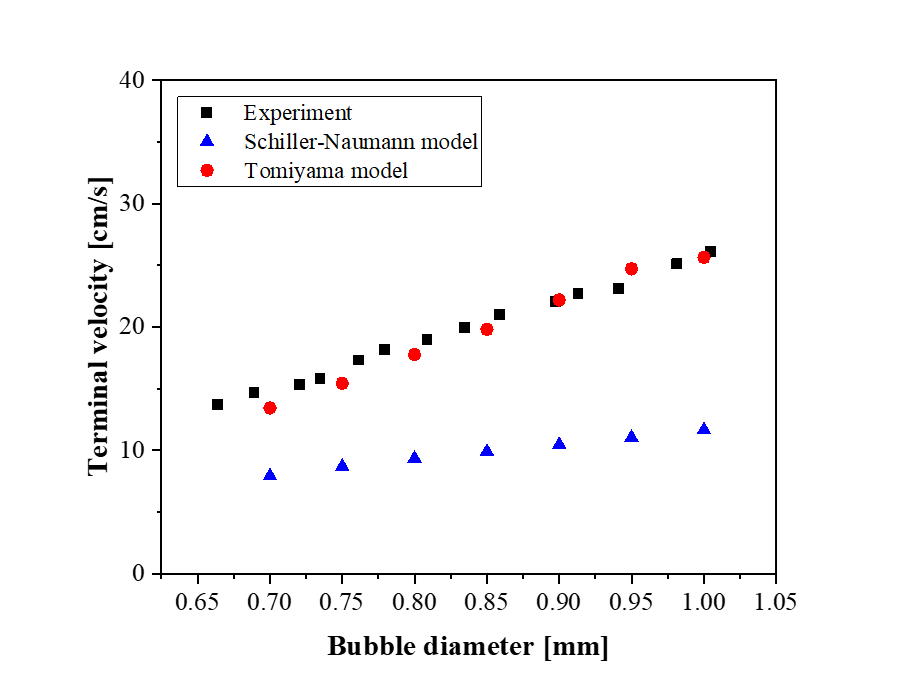}
    \caption{Validation of various models' bubble rising velocity results according to bubble diameter}
    \label{fig:Terminal velocity validation}
\end{figure}
The bubble rising velocity is an important factor in the bubble drag reduction achieved as it determines the bubble location after collision with the wall. Therefore, a suitable bubble drag model leads to more accurate bubble drag reduction simulation results. The Schiller-Naumann model is the traditional particle drag model for modeling rigid spherical particles. While the Tomiyama drag model is an alternative empirical model for simulating spherical bubble. In this section, these two particle drag models are validated using experimental results from previous research\cite{duineveld1995rise}.
Figure. \ref{fig:Terminal velocity validation} shows the validation of these two models by comparing their bubble rising velocity according to bubble diameter with experimental results. The bubbles were located in calm water and their terminal velocities were also calculated. When the bubble floats upwards while the hydrodynamic forces are balanced, the terminal rising velocity is reached. As can be seen in Fig. \ref{fig:Terminal velocity validation}, the Schiller-Naumann model underestimates the rising velocity compared to experimental results. However, the Tomiyama drag model results are in good agreement with the experimental results\cite{duineveld1995rise}. Therefore, the accuracy of the chosen drag model for simulating the bubble rising dynamics is proven. 

\subsubsection{Microbubble dynamic model}
\begin{figure}
    \centering
    \includegraphics[width=1.0\linewidth]{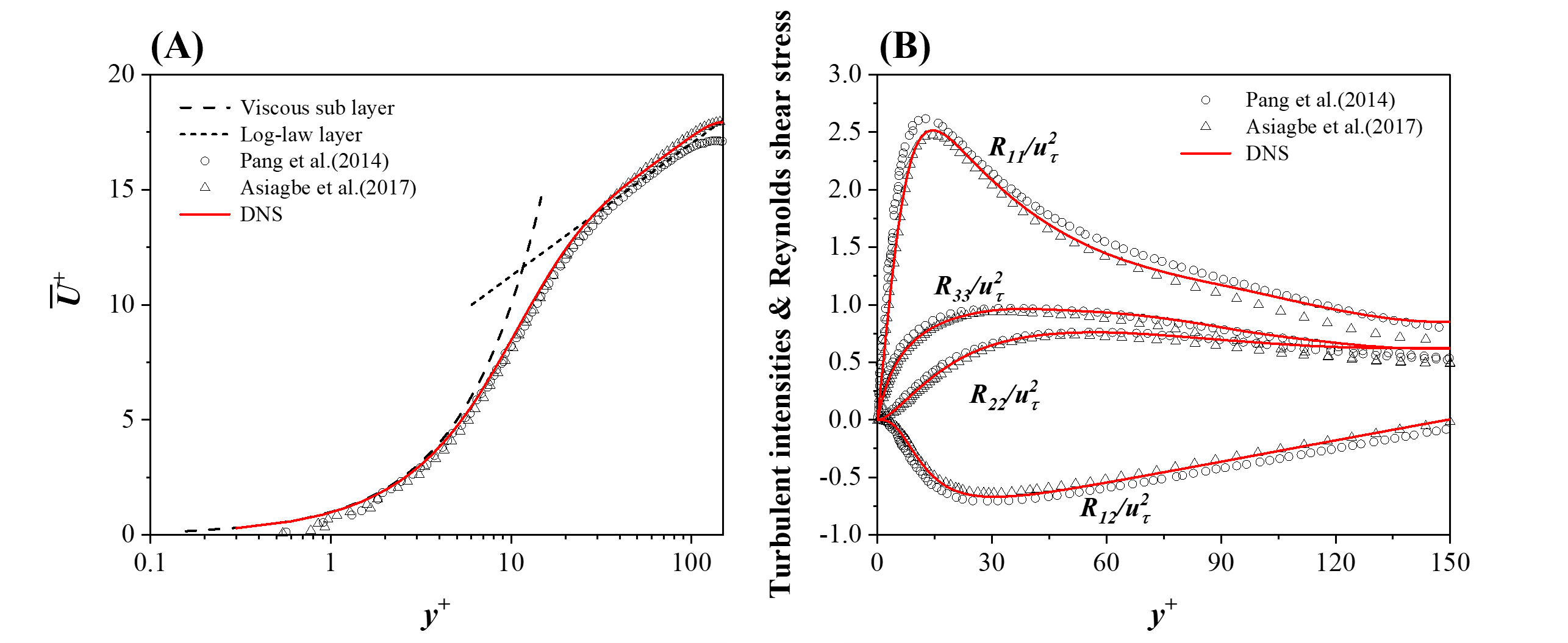}
    \caption{The validation of microbubble tracking code with previous reference\cite{asiagbe2017large,pang2014numerical}}
    \label{fig:Lagrange_code_validation}
\end{figure}

Before carrying out the microbubble-laden turbulent channel flow simulation with SHS, the Lagrange microbubble tracking code and SHS simulations should be validated. First, the microbubble-laden turbulent flow was validated for $Re_{\tau}=150$. Previous research on microbubbles in a horizontal channel flow was conducted using the Euler-Lagrange approach coupled with DNS\cite{pang2014numerical} and LES\cite{asiagbe2017large}. The void fraction of microbubbles, $\alpha$ is 0.0112$\%$, while the size of microbubbles is 110 $\mu m$. Figure. \ref{fig:Lagrange_code_validation} shows a comparison of the calculated velocity profile, turbulent intensities, and Reynolds shear stress profile with previous research. As can be seen in the graphs, the DNS results showed good agreement with previous research\cite{pang2014numerical, asiagbe2017large}. The drag reduction effect from microbubbles was found to be 2.07$\%$, this is similar to a result of 2.12$\%$ from previous research. These results were validated in previous work by Kim et al.\cite{kim2025uncertainty}. 
\begin{figure}[htbp]
    \centering
    \vspace{-2pt} 
    \includegraphics[width=0.85\textwidth]{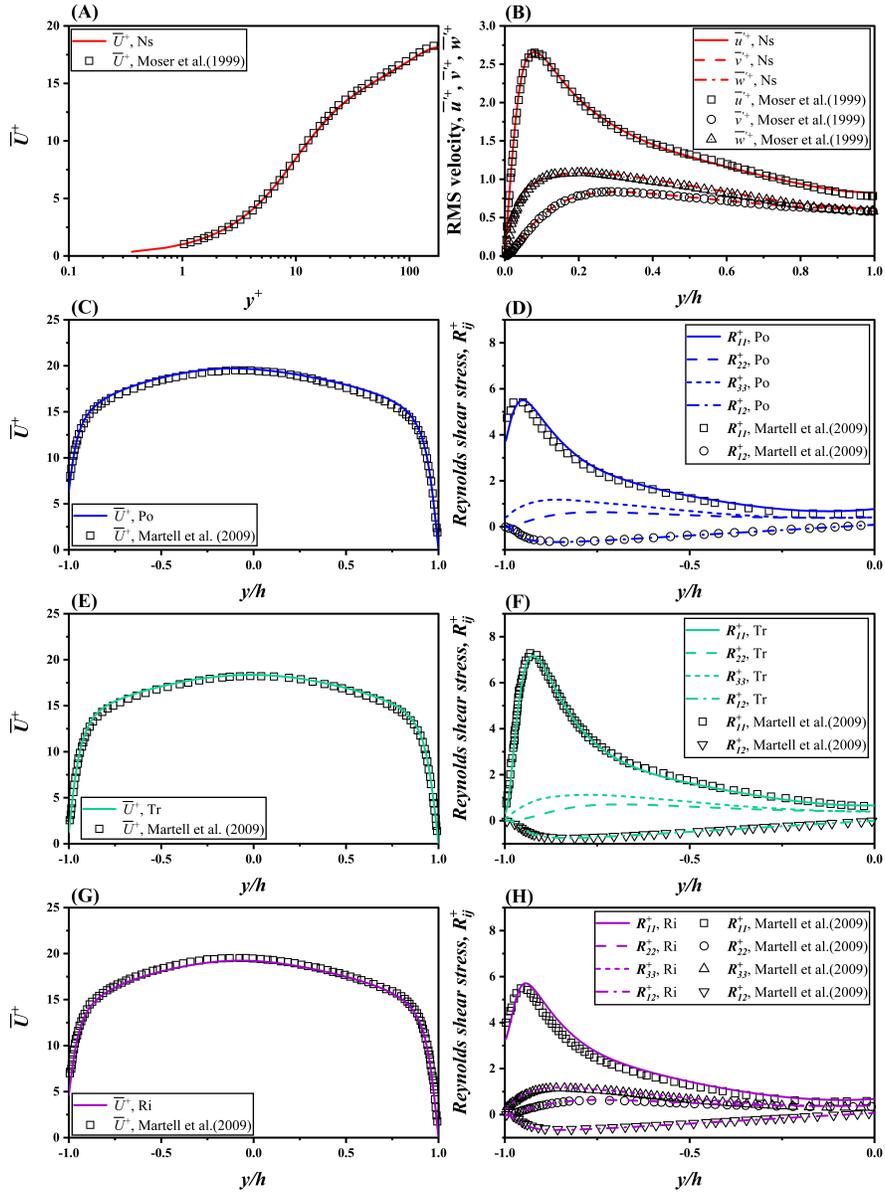}
    \caption{Validation results for different SHS types and a no-slip wall in horizontal channel flow ($Re_\tau=180$).}
    \vspace{-5pt} 
    \label{fig:Validation_lor_SHS}
\end{figure}

\subsubsection{Superhydrophobic surface}
The three types of SHS were simulated and validated against previous studies \cite{martell2009direct,martell2010analysis}. The geometries considered include post type, transverse ridge type, and ridge type, each geometry shares the same width-to-gap ratio ($w/g$) of 1.0. The streamwise velocity profiles for each SHS type are presented in Figure \ref{fig:Validation_lor_SHS}. Panels (A, C, E, G) display the time-averaged streamwise velocity profiles corresponding to the no-slip wall and the three SHS configurations. These profiles show good agreement with the reference data, confirming the validity of the numerical approach. The Reynolds stresses($R_{ij}$) for the same wall conditions are shown in panels (B, D, F, H). The Reynolds stresses are calculated as $R_{ij}=\overline{(u_i-\overline{u_i})(u_j-\overline{u_j})}$. The trends observed in the DNS results closely follow the results from previous research \cite{moser1999direct, martell2009direct, martell2010analysis}, further demonstrating the reliability of the simulation methodology. Overall, the agreement between the present DNS results and the reference data supports the accuracy of the chosen SHS geometries and their implementation in the computational model.

\subsection{Velocity profile}
\label{sec:Velocity profile}
\begin{figure}
    \centering
    \includegraphics[width=1\linewidth]{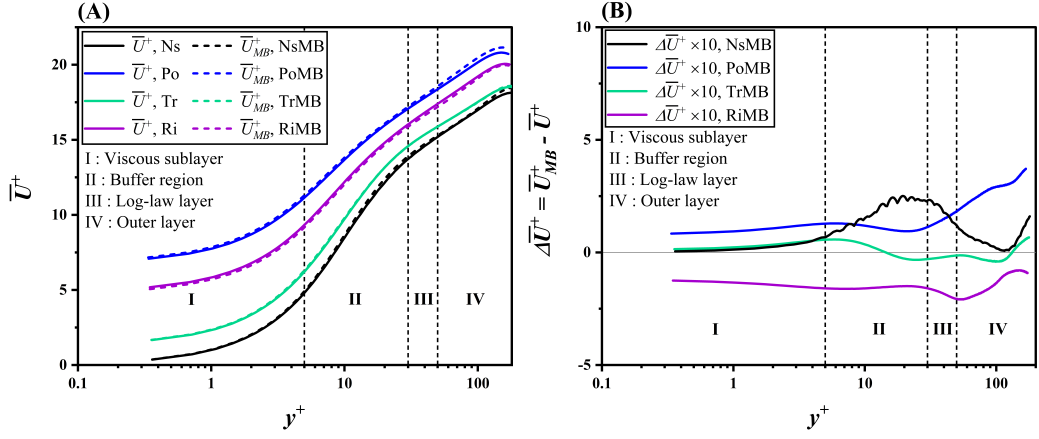}
    \caption{Velocity profiles depending on the presence of microbubbles for different types of wall conditions. (A) Comparison of streamwise velocity profiles, (B) Velocity difference between microbubble-laden and unladen cases.}
    \label{fig:Velocity profile}
\end{figure}
Figure. \ref{fig:Velocity profile} illustrates the non-dimensional streamwise velocity profiles and velocity differences depending on the presence of microbubbles for various wall conditions. The comparison is made for different wall conditions (no-slip wall, post-type SHS, transverse ridge-type SHS, and ridge-type SHS) with and without microbubbles to analyze the flow characteristics.
Figure. \ref{fig:Velocity profile} (A) shows the non-dimensional streamwise velocity($\overline{U}^+$) plotted against the non-dimensional wall-normal distance($y^+$). The results are presented for 4 different wall conditions: no-slip wall(Ns), post-type SHS(Po), transverse ridge-type SHS(Tr), and ridge-type SHS(Ri), for all microbubble-laden cases(shown by dashed lines) and unladen cases(shown by solid lines). For all wall conditions, the presence of microbubbles generally influences the flow velocity in the buffer region (II). The no-slip wall (Ns), post-type SHS (Po), and transverse-ridge-type SHS (Tr) all show a slight increase in velocity when microbubbles are injected. However, in the case of the ridge-type SHS, a slight reduction in velocity is observed in the presence of microbubbles.
Figure. \ref{fig:Velocity profile} (B) presents the velocity difference ($\Delta \overline{U}^+$) between the microbubble-laden flow velocity($\overline{U}^+_{MB}$) and the unladen flow velocity($\overline{U}^+$) for each wall condition. The velocity difference is most notable in the buffer region(II) for the no-slip wall and transverse-ridge-type SHS. The post-type SHS shows the largest positive velocity increments in the outer layer(IV), indicating the microbubbles have a significant effect in these conditions. On the other hand, the ridge-type SHS shows a reduction in velocity with microbubble injection, suggesting an increase in drag when microbubbles are present.
These results demonstrate that the effect of microbubbles varies depending on the wall conditions, highlighting the critical role the SHS patterns play in determining how the interaction between the wall and microbubbles affects drag. Specifically, the post-type SHS appears to be the most effective configuration for reducing drag in combination with microbubbles. Interestingly, the combination of SHS and microbubbles can either increase or decrease drag depending on the SHS pattern used.

\subsection{Drag reduction}
\label{sec:Drag reduction}

\begin{table}
\caption{Comparison of drag reduction for different wall types with and without microbubbles}
\centering\resizebox{\textwidth}{!}{
\begin{tabular}{c c c c c}
 \hline
 Wall type & DR w/o bubble (\%) & DR w/ bubble (\%) & Improvement \\ \hline
 No-slip wall & 0.00 & 1.44 & 1.44 \\
 Post type & 9.32 & 12.57 & 3.25 \\
 Transverse ridge type & 1.12 & 2.50 & 1.38 \\
 Ridge type & 8.68 & 7.54 & -1.14 \\ \hline
\end{tabular}}
\label{table:DR table}
\end{table}

 The drag reduction results from the no-slip wall condition and the three different SHS conditions are compared. Drag reduction is calculated using the equation below :
\begin{equation}
    DR=(1-\frac{\tau_w}{\tau_{w,0}})\times 100
\end{equation}
$\tau_w$ gives the wall shear stress on the wall or SHS. $\tau_{w,0}$ represents the wall shear stress of the no-slip wall for $Re_\tau=180$. The drag reduction results for the 8 cases (4 wall types, with and without bubbles) are shown in Table. \ref{table:DR table}. Looking at this table, we can compare drag reduction values in the presence of microbubbles where for the no-slip wall, drag reduction increases from $0\%$ to $1.44\%$.
For the SHS conditions, we can see the drag reduction performance without microbubbles depends on the patterns of SHS used. The more interesting point is whether microbubble injections causes a drag reduction effect depends on the SHS pattern used. The most effective SHS pattern is the post type, this conditions's drag reduction effect is enhanced from $9.32\%$ to $12.57\%$ in the presence of microbubbles. The transverse ridge type SHS saw its drag reduction performance (relative to the no-slip wall without microbubbles) improve from $1.12\%$ to $2.50\%$ when microbubbles were used. However, the microbubbles cause an increase in drag for the ridge-type SHS as its drag reduction performance goes from $8.68\%$ to $7.54\%$ when microbubbles are introduced.

\begin{table}
\caption{Comparison of slip length,$l_{Slip}^+$ for different SHS types with and without microbubbles}
\centering\resizebox{\textwidth}{!}{
\begin{tabular}{c c c c c}
 \hline
 SHS type & DR w/o bubble & DR w/ bubble & Improvement \\ \hline
 No-slip wall & 0.00 & $\approx$0.00 & - \\
 Post type & 6.66 & 6.82 & 0.16 \\
 Transverse ridge type & 1.30 & 1.31 & 0.01 \\
 Ridge type & 4.86 & 4.72 & -0.14 \\ \hline
\end{tabular}}
\label{table:slip length table}
\end{table}

Table. \ref{table:slip length table} presents a comparison of slip length($l_{Slip}^+$) for different wall types with and without microbubbles. The slip length, originally proposed by Navier\cite{navier1822memoire} is calculated as follows:
\begin{equation}
    u_{x, Slip}=l_{slip}{\frac{\partial u_x}{\partial y}}\Bigr|_{B}
\end{equation}
Here, $u_{x,Slip}$ represents the slip velocity, which corresponds to the streamwise velocity at the SHS surface (i.e., the bottom surface). ${\frac{\partial u_x}{\partial y}}\Bigr|_{B}$ denotes the streamwise velocity gradient at the wall. Slip length is widely recognized as an effective indicator of drag reduction, as several studies have demonstrated its strong correlation with drag reduction performance\cite{park2013numerical,choi2006large,maynes2007laminar,jung2009biomimetic}. In particular, a larger slip length generally corresponds to greater drag reduction\cite{min2004effects,park2013numerical,bidkar2014skin}. Table. \ref{table:slip length table} reveals that the introduction of microbubbles enhances the slip length for post-type SHS and transverse-ridge-type SHS, thereby improving drag reduction. However, in the case of the ridge-type SHS with microbubbles, the slip length decreases, which indicates a weakening of the drag reduction effect due to altered flow behavior near that surface.

\subsection{Turbulent profiles}
\label{sec:Turbulent profiles}
\begin{figure}
    \centering
    \includegraphics[width=1\linewidth]{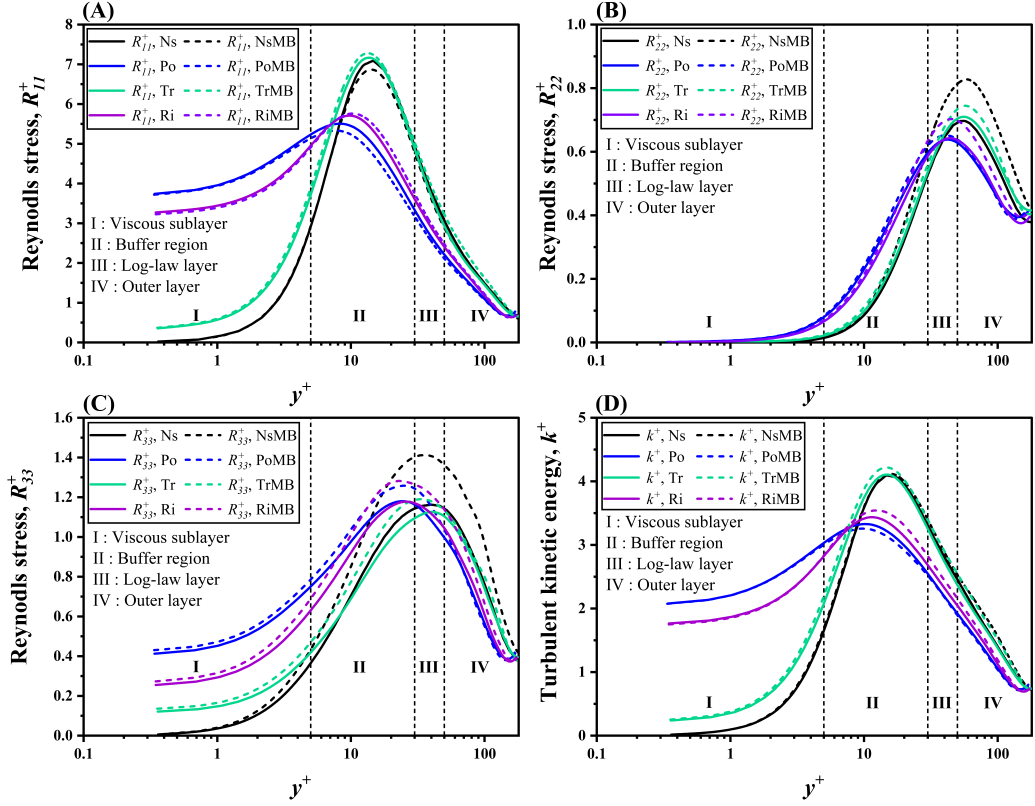}
    \caption{Turbulent kinetic energy profile according to wall type, with and without microbubbles. \\
             (A) No-slip wall, (B) Post-type SHS, (C) Transverse-ridge-type SHS, (D) Ridge-type SHS }
    \label{fig:TKE profile}
\end{figure}

Figure. \ref{fig:TKE profile} shows the distribution of Reynolds stresses($R_{ii}^+$) and turbulent kinetic energy ($k^+=\frac{1}{2} (R_{11}^+ +R_{22}^+ +R_{33}^+)$) over the wall-normal direction ($y^+$) for various wall conditions. Figure. \ref{fig:TKE profile} (A) shows the streamwise Reynolds stress component. Microbubble injection over the no-slip wall (NsMB) leads to attenuation of the peak $R_{11}^+$ value in the buffer region (II) compared to the same wall without bubbles (Ns). This reduction highlights the ability of microbubbles to suppress turbulence intensity. This is a similar trend to previous studies\cite{asiagbe2017large, kim2025uncertainty}. Among the SHS configurations, the post-type with microbubbles (PoMB) exhibits the most noticeable reduction in $R_{11}^+$ compared to the baseline (Ns), indicating this case's strong turbulence-dampening effect. Interestingly, however, microbubbles injection with the other SHS configuration (RiMB, TrMB) leads to a slight increase in the $R_{11}^+$ peak compared to the same walls without microbubbles (Ri, Tr). This opposite trend implies that for the ridge-type and transverse ridge-type SHSs, microbubbles interact differently with the turbulent flow, potentially enhancing rather than suppressing the streamwise Reynolds stress. In (B) and (C), the wall-normal ($R_{22}^+$) and spanwise ($R_{33}^+$) Reynolds stress components show variations in magnitude, both increase in the presence of microbubbles. The magnitude of this increase varies according to wall type.

The NsMB case had the largest increase in $R_{22}^+$ and $R_{33}^+$ due to the presence of microbubbles, with maximum increases of 22.4\% and 22.7\%, respectively. Among the SHS cases, the RiMB cases saw maximum increases of 9.6\% and 10.8\% in $R_{22}^+$ and $R_{33}^+$. Additionally, the PoMB case had maximum increases of 4.1\% in $R_{22}^+$ and 7.1\% in $R_{33}^+$ with microbubble injection. Finally, the TrMB case exhibited maximum increases of 5.9\% and 10.6\% in $R_{22}^+$ and $R_{33}^+$, respectively, with microbubble injection. In all conditions, the maximum increase due to the presence of microbubbles in $R_{33}^+$ was greater than that in $R_{22}^+$. In both panels, a consistent trend can be observed: microbubble injection enhances turbulence intensity in these Reynolds stress components rather than suppressing them. This trend is most pronounced, ie., these increases in Reynolds stress are most significant, in the buffer region for $R_{22}^+$ and in the log-law layer for $R_{33}^+$. 

Panel (D) shows the turbulent kinetic energy (TKE) profiles for the various cases. The shape of these TKE profiles are similar to the $R_{11}^+$ profiles. Among the SHS configurations, only the post-type SHS exhibits a reduction in TKE due to the presence of microbubbles, while the other two SHS configurations show an increase. The no-slip wall case shows slight increase in TKE with the injection of microbubbles over the surface. The changes induced by microbubbles are most pronounced in the buffer region, where clear differences can be observed.


\subsection{Coherent structure}
\label{sec:Coherent structure}

\begin{landscape}
\thispagestyle{mylandscape}
\begin{figure}[htbp]
    \centering
    \vspace{-10pt} 
    \centerline{\includegraphics[width=1.3\textwidth, keepaspectratio]    {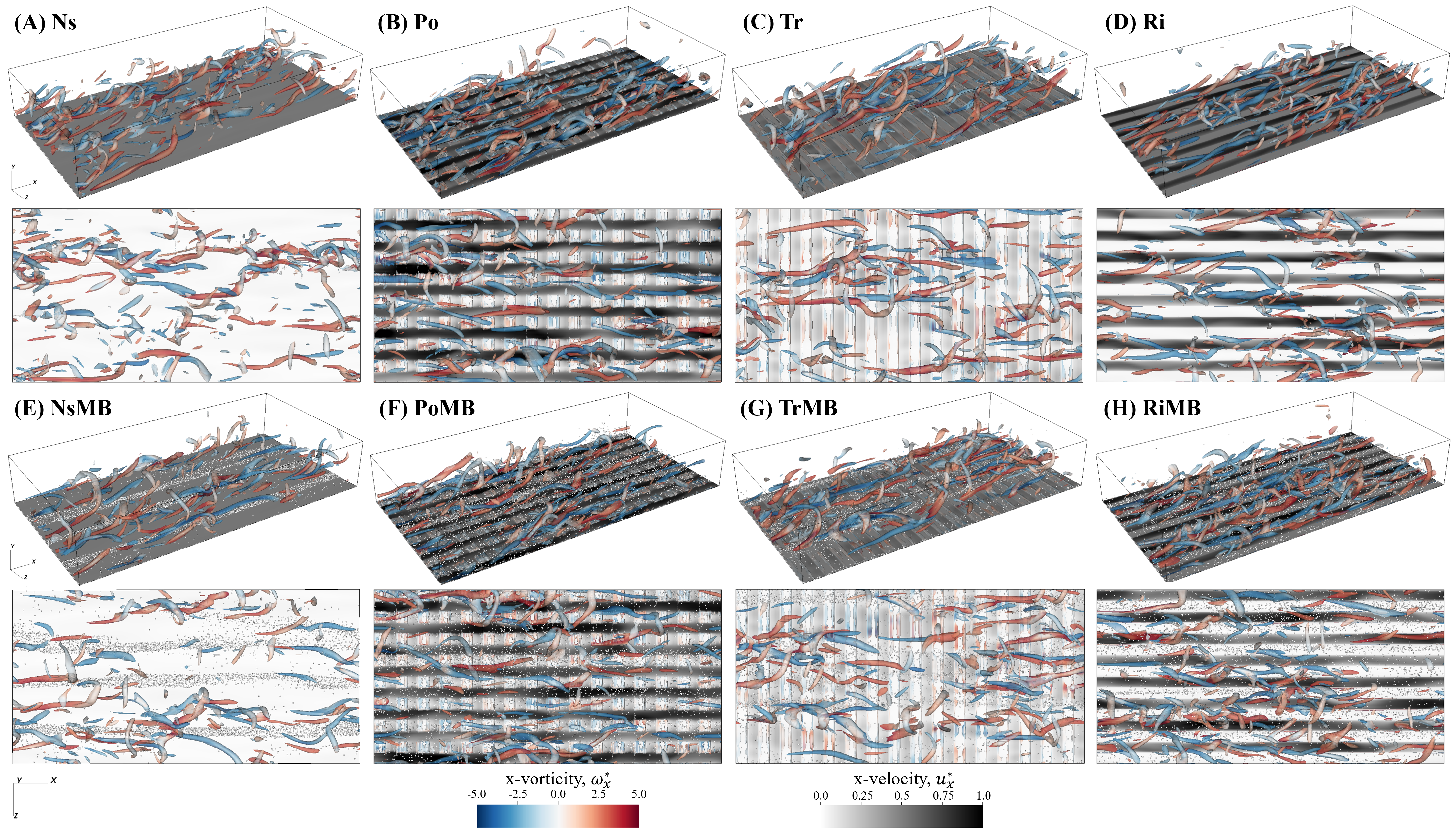}}
    \vspace{-5pt} 
    \caption{Visualization of coherent structures and flow fields for various wall conditions, using streamwise vorticity ($\omega_x^*$) and streamwise velocity ($u_x^*$) contours. Panels (A)-(D) represent cases without microbubbles: (A) No-slip wall, (B) Post-type SHS, (C) Transverse ridge-type SHS, and (D) Ridge-type SHS. Panels (E)-(H) include microbubbles: (E) No-slip wall with microbubbles, (F) Post-type SHS with microbubbles, (G) Transverse ridge-type SHS with microbubbles, and (H) Ridge-type SHS with microbubbles. Blue and red colors indicate negative and positive x-vorticity, respectively, while grayscale contours represent the streamwise velocity. The effect of microbubbles and the various surface geometries on the coherent structures and turbulence modulation is evident, particularly in the buffer region near the wall.}
    \label{fig:Coherent structure}
\end{figure}
\end{landscape}

Figure. \ref{fig:Coherent structure} visualizes the coherent structures and flow fields for various wall conditions with and without microbubbles in both 3D and top-down views. The coherent structures are visualized using the streamwise vorticity component($\omega_x^*$) and overlaid with contours of the streamwise velocity($u_x^*$) at $y^+\simeq0.4$, where blue and red denote negative and positive vorticity, respectively. Each panel highlights the effects of various wall types and of microbubbles injection on the flow structures. Panels (A-D) correspond to the cases without microbubbles: no-slip wall, post-type SHS, transverse-ridge-type SHS, and ridge-type SHS. The coherent structures were extracted using Q-criterion\cite{dubief2000coherent}. Panels (E-H) show the coherent structures and flow fields for the same wall types, this time with microbubble injection. Microbubbles are shown at double size for clarity and convenience.
The first thing we notice is how the hair-pin vortices are mitigated by microbubbles injection near the outer layer. In the NsMB (E) and PoMB (F) cases, the quantity and intensity of the coherent structures are significantly reduced after microbubble injection, indicating microbubbles are able to effectively suppress turbulence. Conversely, in the RiMB (H) and TrMB (G) cases, the coherent structure density remains relatively stable, or even slightly increases in some regions after microbubble injection. This observation implies that the ridge-type SHS and transverse-ridge-type SHS have limited turbulence suppression when combined with microbubbles, highlighting the variation in how well these two drag reduction approaches work together depending on the SHS pattern used. The common features of microbubble injection, regardless of wall type, is that the coherent structures are reduced in size in their presence.

Panel (E) shows the microbubble distribution for the no-slip wall case. In this case, as reported in previous research\cite{zhai2020simulation,mattson2011simulation,zhang2020euler,park2018bubbly}, filament-like strips can be observed. Streamwise vortical structures induce the microbubbles to concentrate preferentially in the low-speed streak. However, the microbubble distributions show different trends in the SHS cases. For the ridge-type SHS, the microbubbles are preferentially concentrated in the no-slip zones. Relative to (E)NsMB, we can see more widely dispersed-filament-like strips of microbubbles in (G)TrMB. However, these filament-like microbubble strips are not seen in (F)PoMB, instead the microbubbles are more evenly and widely distributed in the flows.


\begin{landscape}
\thispagestyle{mylandscape}
\begin{figure}[ht!]
    \centering
    \centerline{\includegraphics[width=1.5\textwidth]{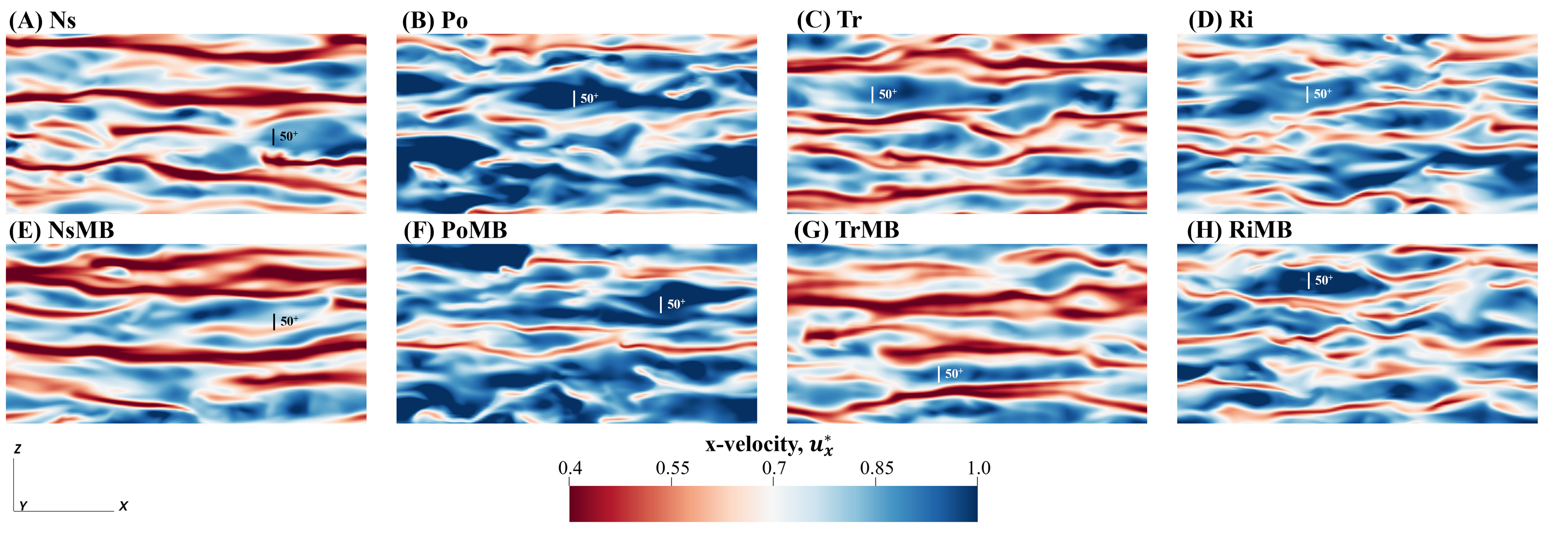}}
    \caption{Instantaneous streamwise velocity contours for various wall types at $y^+\simeq 15$.}
    \label{fig:velocity contour}
\end{figure}
\end{landscape}

Figure. \ref{fig:velocity contour} shows the instantaneous streamwise velocity ($u_x^*$) contours for the various wall types, both with and without microbubbles, at $y^+\simeq15$. Panels (A)-(D) show cases without microbubbles, while panels (E)-(H) show cases with microbubble injection. The red and blue contours indicate low and high streamwise velocity regions, respectively, highlighting the flow field characteristics. Streaks, characterized by pairs of counter-rotating vortices, exhibit an average spanwise spacing of approximately $100^+$ units\cite{kim1987turbulence}. The panels all include a reference bar representing $50^+$ wall units to help compare the relative scales of flow structures between cases.
The first thing you see when looking at these panels are the broad streaks in the Ns and Tr cases, these are related to the filament-like strip structures of microbubbles driven by a broad low-speed streak. In contrast, Po and Ri have narrower, shorter low-speed streaks. This is likely due to the shifted velocity profile caused by the SHS condition. As noted in a previous study by Martell et al.\cite{martell2010analysis}, the velocity distribution is shifted away from the wall by the SHS, similar turbulent structures appear closer to the wall than in the no-slip wall condition. A common effect seen in all cases with microbubbles injection is an increase in the length of the low-speed streaks. In the NsMB and PoMB cases, the low-speed streaks are widened by microbubble injection compared to the Ns and Po cases. In the TrMB and RiMB cases, the opposite happens, the gaps between streaks are narrowed with microbubble injection. Looking at Figure. \ref{fig:TKE profile}, we can see this behavior is consistent with how the TKE($k^+$) profiles are affected by microbubble injection. How the spacing of velocity streaks are affected by microbubbles is also mentioned in the work of Xu et al.\cite{xu2002numerical}. It is interesting to note that the spacing of the streaks decreases in the Ri and Tr cases with microbubble injection, while we see the opposite effect with microbubble injection along a normal wall.

\section{Conclusion}
\label{sec:Conclusion}
A DNS of turbulent channel flow with microbubble injection over various SHS types was conducted using a two-way coupling Euler-Lagrange approach and the SEM. For horizontal channel with $Re_{\tau}=180$, we investigated synergistic drag reduction by combining microbubble injection and SHS, various SHS patterns were studied. Four wall conditions were considered: no-slip wall, post-type SHS, transverse-ridge-type SHS, and ridge-type SHS. As has been shown in a wide array of previous research, applying a single drag reduction method in isolation gives a clear improvement in performance. However, combining two different drag reduction approaches does not always result in a straightforward improvement in performance. The key findings from our study are as follows: \newline

1. The presence of microbubbles enhances streamwise velocity over post-type SHS, transverse-ridge-type SHS, and no-slip walls, whereas microbubble injection over a ridge-type SHS leads to a reduction in streamwise velocity.

2. The ridge-type SHS experiences a decline in performance with microbubbles injection, going from 8.67$\%$ drag reduction (Ri) relative to a no-slip wall with no microbubbles (Ns) to 7.54$\%$ drag reduction (RiMB). Conversely, post-type SHS, transverse-ridge-type SHS, and no-slip walls all exhibit enhanced synergistic drag reduction effects with microbubble injection.

3. Microbubble injection reduces the streamwise Reynolds stress ($R_{11}^+$) for the no-slip wall conditions. However, transverse-ridge-type SHS and ridge-type SHS both see increases in $R_{11}^+$ with microbubble injection. Similarly, turbulent kinetic energy ($k^+$) also increases with microbubble injection in these configurations. In contrast, microbubble injection diminishes both the streamwise Reynolds stress and turbulent kinetic energy in the post-type SHS case.

4. In the no-slip wall case, filament-like strips of microbubbles appear, while the distribution of microbubbles varies with wall surface characteristics in the other cases. In the transverse-ridge-type SHS case, widely distributed filament-like strips of microbubbles were observed, while in the post-type SHS case, the microbubbles were more evenly distributed. In contrast, microbubbles predominantly accumulate in no-slip regions of the ridge-type SHS case. These findings suggest that wall surface characteristics can be leveraged to control microbubble behavior.




%

%
%

\section{Acknowledgment}
\label{sec:Acknowledgment}
The research was supported by the Agency for Defense Development by the Korean Government (UD230502DD).

 \bibliographystyle{elsarticle-num} 
 \bibliography{cas-refs}

\begin{thebibliography}{10}
\expandafter\ifx\csname url\endcsname\relax
  \def\url#1{\texttt{#1}}\fi
\expandafter\ifx\csname urlprefix\endcsname\relax\def\urlprefix{URL }\fi
\expandafter\ifx\csname href\endcsname\relax
  \def\href#1#2{#2} \def\path#1{#1}\fi

\bibitem{fukuda2000frictional}
K.~Fukuda, J.~Tokunaga, T.~Nobunaga, T.~Nakatani, T.~Iwasaki, Y.~Kunitake, Frictional drag reduction with air lubricant over a super-water-repellent surface, Journal of Marine Science and Technology 5 (2000) 123--130.

\bibitem{wang2022drag}
H.~Wang, K.~Wang, G.~Liu, Drag reduction by gas lubrication with bubbles, Ocean Engineering 258 (2022) 111833.

\bibitem{white2008mechanics}
C.~M. White, M.~G. Mungal, Mechanics and prediction of turbulent drag reduction with polymer additives, Annu. Rev. Fluid Mech. 40~(1) (2008) 235--256.

\bibitem{marchioli2021drag}
C.~Marchioli, M.~Campolo, Drag reduction in turbulent flows by polymer and fiber additives, KONA Powder and Particle Journal 38 (2021) 64--81.

\bibitem{ceccio2010friction}
S.~L. Ceccio, Friction drag reduction of external flows with bubble and gas injection, Annual Review of Fluid Mechanics 42 (2010) 183--203.

\bibitem{murai2014frictional}
Y.~Murai, Frictional drag reduction by bubble injection, Experiments in Fluids 55 (2014) 1--28.

\bibitem{rothstein2010slip}
J.~P. Rothstein, Slip on superhydrophobic surfaces, Annual review of fluid mechanics 42~(1) (2010) 89--109.

\bibitem{lee2016superhydrophobic}
C.~Lee, C.-H. Choi, C.-J. Kim, Superhydrophobic drag reduction in laminar flows: a critical review, Experiments in Fluids 57 (2016) 1--20.

\bibitem{park2021superhydrophobic}
H.~Park, C.-H. Choi, C.-J. Kim, Superhydrophobic drag reduction in turbulent flows: A critical review, Experiments in Fluids 62 (2021) 1--29.

\bibitem{dean2010shark}
B.~Dean, B.~Bhushan, Shark-skin surfaces for fluid-drag reduction in turbulent flow: a review, Philosophical Transactions of the Royal Society A: Mathematical, Physical and Engineering Sciences 368~(1929) (2010) 4775--4806.

\bibitem{jang2014experimental}
J.~Jang, S.~H. Choi, S.-M. Ahn, B.~Kim, J.~S. Seo, Experimental investigation of frictional resistance reduction with air layer on the hull bottom of a ship, International Journal of Naval Architecture and Ocean Engineering 6~(2) (2014) 363--379.

\bibitem{tanaka2021repetitive}
T.~Tanaka, Y.~Oishi, H.~J. Park, Y.~Tasaka, Y.~Murai, C.~Kawakita, Repetitive bubble injection promoting frictional drag reduction in high-speed horizontal turbulent channel flows, Ocean Engineering 239 (2021) 109909.

\bibitem{breveleri2023plastron}
J.~Breveleri, S.~Mohammadshahi, T.~Dunigan, H.~Ling, Plastron restoration for underwater superhydrophobic surface by porous material and gas injection, Colloids and Surfaces A: Physicochemical and Engineering Aspects 676 (2023) 132319.

\bibitem{zhu2024achieving}
D.~Zhu, Y.~Song, F.~Gao, S.~Dong, C.~Xu, B.~Liu, J.~Zheng, X.~Zhou, Q.~Liu, Achieving underwater stable drag reduction on superhydrophobic porous steel via active injection of small amounts of air, Ocean Engineering 308 (2024) 118329.

\bibitem{fischer2008nek5000}
P.~F. Fischer, J.~W. Lottes, S.~G. Kerkemeier, et~al., nek5000 web page (2008).

\bibitem{patera1984spectral}
A.~T. Patera, A spectral element method for fluid dynamics: laminar flow in a channel expansion, Journal of computational Physics 54~(3) (1984) 468--488.

\bibitem{asiagbe2017large}
K.~S. Asiagbe, M.~Fairweather, D.~O. Njobuenwu, M.~Colombo, Large eddy simulation of microbubble transport in a turbulent horizontal channel flow, International Journal of Multiphase Flow 94 (2017) 80--93.

\bibitem{elgobashi2006updated}
S.~Elgobashi, An updated classification map of particle-laden turbulent flows, in: IUTAM Symposium on Computational Approaches to Multiphase Flow: Proceedings of an IUTAM Symposium held at Argonne National Laboratory, October 4--7, 2004, Springer, 2006, pp. 3--10.

\bibitem{pang2014numerical}
M.~Pang, J.~Wei, B.~Yu, Numerical study on modulation of microbubbles on turbulence frictional drag in a horizontal channel, Ocean Engineering 81 (2014) 58--68.

\bibitem{velasco2022numerical}
L.~J. Velasco, D.~N. Venturi, D.~H. Fontes, F.~J. de~Souza, Numerical simulation of drag reduction by microbubbles in a vertical channel, European Journal of Mechanics-B/Fluids 92 (2022) 215--225.

\bibitem{zhai2020simulation}
J.~Zhai, M.~Fairweather, M.~Colombo, Simulation of microbubble dynamics in turbulent channel flows, Flow, Turbulence and Combustion 105 (2020) 1303--1324.

\bibitem{maxey1983equation}
M.~R. Maxey, J.~J. Riley, Equation of motion for a small rigid sphere in a nonuniform flow, The Physics of Fluids 26~(4) (1983) 883--889.

\bibitem{schiller1933drag}
L.~Schiller, A drag coefficient correlation, Zeit. Ver. Deutsch. Ing. 77 (1933) 318--320.

\bibitem{tomiyama2002terminal}
A.~Tomiyama, G.~Celata, S.~Hosokawa, S.~Yoshida, Terminal velocity of single bubbles in surface tension force dominant regime, International journal of multiphase flow 28~(9) (2002) 1497--1519.

\bibitem{duineveld1995rise}
P.~Duineveld, The rise velocity and shape of bubbles in pure water at high reynolds number, Journal of Fluid Mechanics 292 (1995) 325--332.

\bibitem{brennen1982review}
C.~E. Brennen, A review of added mass and fluid inertial forces, Tech. rep. (Jan 1982).

\bibitem{legendre1998lift}
D.~Legendre, J.~Magnaudet, The lift force on a spherical bubble in a viscous linear shear flow, Journal of Fluid Mechanics 368 (1998) 81--126.

\bibitem{mortimer2019near}
L.~Mortimer, D.~Njobuenwu, M.~Fairweather, Near-wall dynamics of inertial particles in dilute turbulent channel flows, Physics of Fluids 31~(6) (2019) 063302.

\bibitem{min2004effects}
T.~Min, J.~Kim, Effects of hydrophobic surface on skin-friction drag, Physics of Fluids 16~(7) (2004) L55--L58.

\bibitem{min2005effects}
T.~Min, J.~Kim, Effects of hydrophobic surface on stability and transition, Physics of fluids 17~(10) (2005).

\bibitem{martell2009direct}
M.~B. Martell, J.~B. Perot, J.~P. Rothstein, Direct numerical simulations of turbulent flows over superhydrophobic surfaces, Journal of Fluid Mechanics 620 (2009) 31--41.

\bibitem{martell2010analysis}
M.~B. Martell, J.~P. Rothstein, J.~B. Perot, An analysis of superhydrophobic turbulent drag reduction mechanisms using direct numerical simulation, Physics of Fluids 22~(6) (2010).

\bibitem{pope2000turbulent}
S.~B. Pope, S.~B. Pope, Turbulent flows, Cambridge university press, 2000.

\bibitem{grace1976shapes}
J.~R. Grace, Shapes and velocities of single drops and bubbles moving freely through immisicible liquids, Trans. Inst. Chem. Eng. 54 (1976) 167--173.

\bibitem{clift2005bubbles}
R.~Clift, J.~Grace, M.~Weber, \href{https://books.google.co.kr/books?id=UUrOmD8niUQC}{Bubbles, Drops, and Particles}, Dover Civil and Mechanical Engineering Series, Dover Publications, 2005.
\newline\urlprefix\url{https://books.google.co.kr/books?id=UUrOmD8niUQC}

\bibitem{fischer2003implementation}
P.~Fischer, Implementation considerations for the oifs/characteristics approach to convection problems, Argonne National Laboratory (2003).

\bibitem{offermans2017gather}
N.~Offermans, Gather-scatter library in nek5000: Documentation of the gs library developed by james lottes, Tech. rep., Report (2017).

\bibitem{zwick2020scalable}
D.~Zwick, S.~Balachandar, A scalable euler--lagrange approach for multiphase flow simulation on spectral elements, The International Journal of High Performance Computing Applications 34~(3) (2020) 316--339.

\bibitem{kim2025uncertainty}
B.-C. Kim, K.~Chang, S.-W. Lee, J.~Ryu, M.~Kim, J.~Yoon, Uncertainty quantification for the drag reduction of microbubble-laden fluid flow in a horizontal channel, International Journal of Multiphase Flow 182 (2025) 105059.

\bibitem{moser1999direct}
R.~D. Moser, J.~Kim, N.~N. Mansour, Direct numerical simulation of turbulent channel flow up to re$\tau$= 590, Physics of fluids 11~(4) (1999) 943--945.

\bibitem{navier1822memoire}
C.~Navier, M{\'e}moire sur les lois du mouvement des fluides, {\'e}diteur inconnu, 1822.

\bibitem{park2013numerical}
H.~Park, H.~Park, J.~Kim, A numerical study of the effects of superhydrophobic surface on skin-friction drag in turbulent channel flow, Physics of Fluids 25~(11) (2013).

\bibitem{choi2006large}
C.-H. Choi, C.-J. Kim, Large slip of aqueous liquid flow over a nanoengineered superhydrophobic surface, Physical review letters 96~(6) (2006) 066001.

\bibitem{maynes2007laminar}
D.~Maynes, K.~Jeffs, B.~Woolford, B.~Webb, Laminar flow in a microchannel with hydrophobic surface patterned microribs oriented parallel to the flow direction, Physics of fluids 19~(9) (2007).

\bibitem{jung2009biomimetic}
Y.~C. Jung, B.~Bhushan, Biomimetic structures for fluid drag reduction in laminar and turbulent flows, Journal of Physics: Condensed Matter 22~(3) (2009) 035104.

\bibitem{bidkar2014skin}
R.~A. Bidkar, L.~Leblanc, A.~J. Kulkarni, V.~Bahadur, S.~L. Ceccio, M.~Perlin, Skin-friction drag reduction in the turbulent regime using random-textured hydrophobic surfaces, Physics of Fluids 26~(8) (2014).

\bibitem{dubief2000coherent}
Y.~Dubief, F.~Delcayre, On coherent-vortex identification in turbulence, Journal of turbulence 1~(1) (2000) 011.

\bibitem{mattson2011simulation}
M.~Mattson, K.~Mahesh, Simulation of bubble migration in a turbulent boundary layer, Physics of Fluids 23~(4) (2011).

\bibitem{zhang2020euler}
X.~Zhang, J.~Wang, D.~Wan, Euler--lagrange study of bubble drag reduction in turbulent channel flow and boundary layer flow, Physics of Fluids 32~(2) (2020).

\bibitem{park2018bubbly}
H.~J. Park, Y.~Tasaka, Y.~Murai, Bubbly drag reduction accompanied by void wave generation inside turbulent boundary layers, Experiments in fluids 59 (2018) 1--15.

\bibitem{kim1987turbulence}
J.~Kim, P.~Moin, R.~Moser, Turbulence statistics in fully developed channel flow at low reynolds number, Journal of fluid mechanics 177 (1987) 133--166.

\bibitem{xu2002numerical}
J.~Xu, M.~R. Maxey, G.~E. Karniadakis, Numerical simulation of turbulent drag reduction using micro-bubbles, Journal of Fluid Mechanics 468 (2002) 271--281.

\end{thebibliography}





\end{document}